\documentclass[fleqn,12pt,twoside]{article}
\usepackage{espcrc1}
\usepackage{graphicx}

\title{Statistical Fluctuations as Probes of Dense Matter}
\author{Berndt M\"uller\address{Department of Physics, 
        Duke University, Durham, NC 27705, USA}
        \thanks{Work supported in part by the U.S. Department
        of Energy under grant DE-FG02-96ER40945}}

\begin{document}
\maketitle

\begin{abstract}
The use of statistical fluctuations as probes of the microscopic
dynamics of hot and dense hadronic matter is reviewed. Critical
fluctuations near the critical point of QCD matter are predicted
to enhance fluctuations in pionic observables. Chemical fluctuations,
especially those of locally conserved quantum numbers, such as
electric charge and baryon number, can probe the nature of the 
carriers of these quantum numbers in the dense medium.
\end{abstract}

\section{INTRODUCTION}

Thermal models for the production of final-state hadrons in
nuclear collisions have been extremely successful over a wide
range of collision energies ranging from the Bevalac/SIS, over 
the AGS and SPS, to the RHIC \cite{Raf,BMS,Hei,Cley,Flor}.
The parametric dependences of the temperature $T$ and baryon
chemical potential $\mu$ on the nucleon-nucleon center-of-mass
energy $\sqrt{s}$ map out two lines in the phase diagram, which
are usually called the {\em chemical} and {\em thermal freeze-out}
lines. The chemical freeze-out line $(T_{\rm ch},\mu_{\rm ch})$
is deduced from the abundance ratios of different species of hadrons,
while the thermal freeze-out line $(T_{\rm th},\mu_{\rm th})$ is
deduced from the slopes of the transverse momentum spectra of the
emitted hadrons. 

What do these results tell us about QCD except that the hadronic
matter at freeze-out is quite well described by a state of thermal
and chemical equilibrium? Firstly, rather strong arguments have
been presented that the chemical and thermal equilibrium can only
be established by some complex ``prehadronic'' dynamics, which
requires the presence of gluons as dynamical degrees of fredom
\cite{KMR}. Secondly, the thermal equilibrium description
requires the assumption of a strong transverse collective flow
of the matter at freeze-out, which exhibits an increasing degree
of azimuthal anisotropy as the CM energy increases. This effect,
called elliptic flow \cite{Olli}, indicates the presence of a 
strong transverse pressure very soon (1 fm/c) after the begin of 
the nuclear reaction \cite{Kolb}.

As the chemical freeze-out parameters at SPS and RHIC energies
are very close to the expected phase boundary between hadronic
matter and the quark-gluon plasma, it is reasonable to conjecture
that the produced matter was first thermalized in the deconfined
phase and then evolved through the phase boundary into a thermal
gas of hadrons. The crucial question is whether any signatures
remain in hadronic observables that carry information about this
evolution prior to the final freeze-out. 

As I will explain, there are reasons to believe that statistical 
fluctuations \cite{Heisel} in certain hadronic observables may 
preserve information about earlier times, even though their 
averages only tell us about the freeze-out conditions. It is
here where the finite volume and the finite life-time of the
hadronic fireball produced in relativistic heavy-ion collisions
turn out to be advantages rather than deficiencies. The relative
size of fluctuations in thermodynamic observables, as compared
to the average values, decreases inversely with the volume $V$
of the system. And if the system would evolve infintely slowly,
the fluctuations at freeze-out would only reflect the freeze-out
conditions. 

The talk is organized as follows: After a brief review of
statistical fluctuations in parts of an equilibrated system,
I will discuss fluctuations in the momentum spectra of
emitted particles, and later fluctuations in particle
abundances, or chemical fluctuations, where the main emphasis
will be on the fluctuations of conserved quantum numbers.
The talk will conclude with some experimentally relevant
considerations and an outlook.

\section{STATISTICAL FLUCTUATIONS IN A SUBVOLUME}

Consider small subsystem of a large system in thermodynamic
equilibrium at temperature $T=\beta^{-1}$. Let us ask how the 
average energy $\langle E\rangle_V$ contained in the subvolume 
$V$ changes as a function of temperature. A straightforward
calculation yields:
\begin{equation}
\frac{\partial}{\partial\beta} \langle E\rangle_V
= \frac{\partial}{\partial\beta} 
  \frac{{\rm Tr}_V(H e^{-\beta H})}{{\rm Tr}_V(e^{-\beta H})}
= - \langle E^2\rangle_V + \langle E\rangle_V^2 .
\label{eq1}
\end{equation}
As the right-hand side is just the fluctuation of the energy
contained in $V$, we have obtained the important result, that
the energy fluctuations are determined by the heat capacity
of the matter in the subvolume:
\begin{equation}
\langle \Delta E^2 \rangle_V 
= T^2 \frac{\partial\langle E\rangle_V}{\partial T} 
= T^2 C_V 
\label{eq2}
\end{equation}
where $C_V$ is the heat cpaacity.
Fluctuations of other (conserved) quantities $\cal O$ can be
calculated in a similar way, by considering the variation of
the grand canonical average ${\cal O}_{V,T}$ with respect to
a change in the chemical potential $\mu$ associated with $\cal O$:
\begin{equation}
\langle \Delta {\cal O}^2 \rangle_V 
= T \frac{\partial\langle{\cal O}\rangle_{V,T}}{\partial\mu} 
\equiv T \chi_O .
\label{eq3}
\end{equation}
The expression on the right-hand side, apart from the factor $T$, 
is called the {\em susceptibility} $\chi_{\cal O}$, which measures 
the response of the medium to a change in the chemical potenial.
associate with the observable $\cal O$. An important example is the 
magnetic susceptibility $chi_M$, where the observable is given by 
the magnetization $M$, and the magnetic field $H$ assumes the role of 
the chemical potential. Examples of interest in the case of hadronic 
matter are the chiral susceptibility $\chi_m$ and the Polyakov loop
susceptibility $\chi_L$, which measure the response of the medium
to a change in the quark mass, and to the addition of a free heavy
quark, respectively.

\section{THERMAL FLUCTUATIONS}

As outlined above, fluctuations of the thermal energy in a given
subvolume are a measure of the change of the average thermal energy
with temperature, i.e. the heat capacity of the matter contained in
the subvolume. A similar relation connects the fluctuations of the
local temperature to the heat capacity $C_V$:
\begin{equation}
\frac{\langle \Delta T^2 \rangle}{T^2} = \frac{1}{C_V} .
\label{eq4}
\end{equation}
This relation could be used to measure the specific heat of the
hadronic matter created in high-energy nuclear collisions
\cite{St95,Sh98}. The idea
here is to use the slope of the transverse momentum spectrum of
emitted particles as a measure of the temperature -- assuming that
the matter is thermalized -- and to look for fluctuations in the
event ensemble. Near a critical point, the specific heat diverges, 
causing the thermal fluctuations to vanish. A pronounced decrease
in the observed temperature fluctuations between otherwise identical
events would, therefore, indicate an approach to the critical point
in the phase diagram.

The simplest way to identify temperature fluctuations is to look for
fluctuations in the mean transverse momentum $\langle p_T \rangle$.
Since the increase of $C_V$ near $T_c$ is caused by the increased
fluctuations of the modes associated with the order parameter -- in
the case of the chiral phase transition, the quark condensate
$\langle {\bar q}q \rangle$ -- it should be possible to enhance the 
signal by considering event-by-event fluctuations of observables 
most sensitive to these modes. For hadronic matter, this would be 
the mean transverse momentum of low-$p_T$ pions, which are thought 
to partially arise from the decay of local excitations of the 
iso-singlet order parameter, the $\sigma$-meson mode. 

For practical purposes, two observables have been defined \cite{GM92}:
\begin{equation}
\Phi_p = \left( \frac{\langle \Delta P_T^2\rangle}{\langle N\rangle}
         \right)^{1/2} - \left(\overline{\Delta p_T^2}\right)^{1/2}
\label{eq5a}
\end{equation}
where $P_T^2$ is the total squared transverse momentum of all
particles emitted in an event, $N$ is the event multiplicity,
$\langle\cdots\rangle$ denotes the event average, and 
$\overline{\Delta p_T^2}$ the average for the inclusive single 
particle distribution; and \cite{SRS}:
\begin{equation}
F_p = \frac{\langle N\rangle \langle \Delta P_T^2/N \rangle}
         {\overline{\Delta p_T^2}} .
\label{eq5b}
\end{equation}
Being a ratio rather than a difference, $F_p$ may be less sensitive
to collective flow effects than $\Phi_p$.  An estimate for the 
expected size of $F$ due to the fluctuations in the $\sigma$-field 
has can be obtained in the framework of the linear sigma model 
\cite{SRS}:
\begin{equation}
F_\sigma - 1 \approx\, 0.14 
\left( \frac{\xi_\sigma}{6\,\rm fm} \right)^2 ,
\label{eq6}
\end{equation}
where $\xi_\sigma$ is the correlation length of the fluctuations.

\begin{figure}[htb]
\includegraphics[width=10cm]{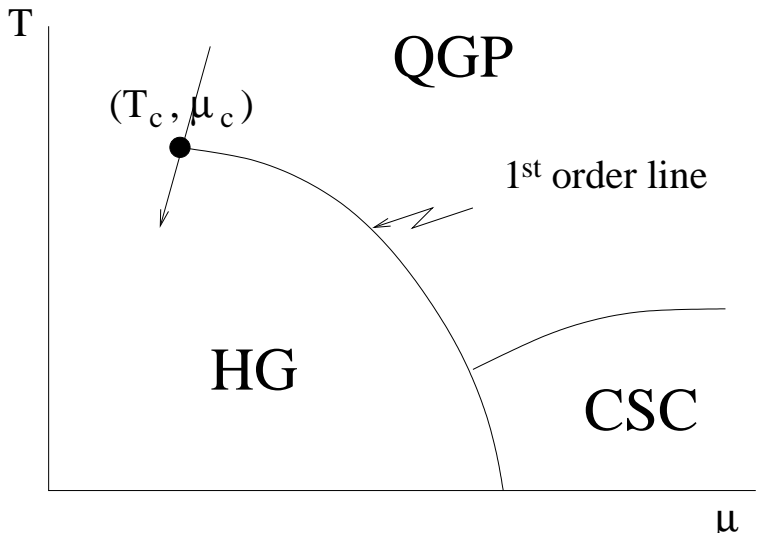}
\caption{Schematic QCD phase diagram. The first-order transition line
separating the quark-gluon plasma (QGP) phase and the hadronic gas (HG) 
phase terminates in a critical point $(T_c,\mu_c)$, where the transition
is of second order. Under favorable conditions, the hot system created
in a heavy ion collision may transiently pass through the critical
point, before it freezes out into hadrons that no longer interact.}
\label{fig1}
\end{figure}

This prediction has to be modified for two reasons. First of all,
the hadronic system does not decay into free hadrons right at the
critical point, even if it reaches this point during its evolution. 
Instead of observing the large correlation length predicted to
occur at $(T_c,\mu_c)$ in equilibrium,
one would thus expect to observe the reduced correlation length
associated with the freeze-out parameters $(T_f,\mu_f)$, see
Fig.~\ref{fig1}. Furthermore, even under conditions of criticality, the
correlation length only diverges if the system has enough time to
develop long-range fluctuations of the order parameter. The theory
of dynamical critical phenomena yields an equation determining the
change in the inverse correlation length $m(t)=\xi(t)^{-1}$ as the
system evolves \cite{BR00}:
\begin{equation}
\frac{dm}{dt} = - \Gamma(m(t))\left(m(t)-m_{\rm eq}(t)\right) .
\label{eq7}
\end{equation}
Here $m_{\rm eq}$ is the inverse correlation length in equilibrium 
-- which vanishes at the critical point -- and the relaxation rate
$\Gamma(m) \sim m^z$ with some positive exponent $z$. As critical
conditions are approached, the relaxation time diverges exhibiting
the well-known effect of critical ``slowing down''. 

If the critical point is reached in a nuclear collision, the system
rapidly expands across it and freezes out soon afterwards. According
to (\ref{eq7}) the correlation length $\xi(t)$ remains smaller than
$\xi_{\rm eq}$ as $T_c$ is approached from above, but then decreases
less rapidly than $\xi_{\rm eq}$. The prediction is thus that the 
effective correlation length at the moment of freeze-out is larger 
than one would expect under equilibrium conditions, reflecting the 
temporary proximity to the critical point \cite{BR00,BDS00}. 

Experimental data for $\langle p_T \rangle$ fluctuations have been 
obtained by NA49 for Pb+Pb collisions at the CERN-SPS \cite{NA49}
and recently by STAR for Au+Au collisions at RHIC \cite{STAR}. 
While the NA49 data reflect mean $p_T$ fluctuations in a forward 
rapidity  region, the STAR data were taken at midrapidity. 
NA49 reported values of $F=1.004\pm 0.004$ or $\Phi_p=0.6\pm 1$ MeV
after corrections for two-particle (HBT) correlations. The value 
from STAR is $\Phi_p\approx 35$ MeV.

\section{CHEMICAL FLUCTUATIONS}

Fluctuations in the chemical composition of the emitted hadron
yields can potentially probe the microscopic structure
of the emitting matter. An example is the quark number
susceptibility 
\begin{equation}
\chi_q = T \frac{\partial}{\partial\mu_q} 
         \langle {\bar q}\gamma^0 q \rangle ,
\label{eq8}
\end{equation}
which measures the response of the net quark density to a change in
the quark chemical potenial. A different, but related quantity is 
the chiral susceptibility $\chi_m = T \partial \langle {\bar q}q 
\rangle / \partial m $, which measures the response of the quark
condensate to a change in the current quark mass. $\chi_q$ has an
isoscalar and an isovector component
\begin{eqnarray}
\chi_{\rm S} & = & T^{-1} \langle 
         [\Delta(u^\dagger u)+\Delta(d^\dagger d)]^2 \rangle ,
\nonumber\\
\chi_{\rm NS} & = & T^{-1} \langle 
         [\Delta(u^\dagger u)-\Delta(d^\dagger d)]^2 \rangle ,
\label{eq9}
\end{eqnarray}
which have both been determined on the lattice \cite{Go87}.
The results show that $\chi_{\rm S}\approx\chi_{\rm NS}$, 
indicating that fluctuations in the u- and d-quark densities
are uncorrelated.

As chemical properties are generally carried by particles,
chemical fluctuations are determined by fluctuations
in the corresponding particle numbers, and their changes are
governed by particle transport processes. A weakly coupled
hadronic gas (HG) and a perturbative quark-gluon plasma (QGP) 
differ significantly in this respect, as listed in Table \ref{tab1}.
\begin{table}[htb]
\caption{Quantum numbers of electric charge ($Q$) and baryon number
         ($B$) of particles and their abundances in a weakly coupled
          hadron gas (HG) and quark-gluon plasma (QGP).} 
\smallskip
\label{tab1}
\begin{tabular}{|l|c|c|c|}
\hline
   & $\vert Q \vert$ & particles & fraction \cr
\hline
   & & & \cr
HG & $e$ & $\pi, K, \ldots$ & $\sim 2/3$ \cr

QGP & $e/3, 2e/3$ & $q, {\bar q}$ & $\sim 1/2$ \cr
\hline
\end{tabular}
\hfill
\begin{tabular}{|l|c|c|c|}
\hline
   & $\vert B\vert$ & particles & fraction \cr
\hline
   & & & \cr
HG & $1$ & $N, {\bar N}, \Lambda, \ldots$ & $\ll 1$ \cr
QGP & $1/3$ & $q, {\bar q}$ & $\sim 1/2$ \cr
\hline
\end{tabular}
\end{table}
This implies that measurable differences in the fluctuations
associated with charge and baryon number exist between a hadron
gas and a quark-gluon plasma. 

The second important point is that fluctuations of locally
conserved quantities, such as the net electric charge or the
net baryon number (or the net strangeness!) cannot be erased
by local reactions. They can only be modified due to particle
transport over larger distances, i.e.~diffusion, and may thus be 
fixated early, if the evolution of the system is sufficiently
rapid \cite{AHM00,JK00}. Consider the total net electric charge
of all particles emitted within a rapidity window $\Delta y$ in 
a high-energy nuclear collision. The fluctuation of this
quantity $\langle \Delta Q^2 \rangle_{\Delta y}$ is proportional
to the volume associated with emission into this rapidity window,
i.e. to $\Delta y$ itself in a boost invariant scenario. On the
other hand, the rate of diffusion of net charge into and out of 
this volume is independent of the size of the rapidity interval.
Therefore, early fluctuations
of the net charge contained in a sufficiently large rapidity
interval $\Delta y \ge \Delta y_{\rm min}$ may survive. The 
questions are, how large is $\Delta y_{\rm min}$, and how much
do the predictions for a HG and a QGP differ?

We first address the last question. We estimate the fluctuations
per unit volume in the two different phases in the weak coupling
limit. In the hadron gas we have
\begin{eqnarray}
\langle \Delta B^2 \rangle_{\rm HG} & = & N_B + N_{\bar B},
\nonumber\\
\langle \Delta Q^2 \rangle_{\rm HG} & \approx & 
  N_{\pi^+} + N_{\pi^-} + N_{K^+} + N_{K^-} + \ldots ,
\label{eq10}
\end{eqnarray}
where $N_i$ denotes the number of hadrons of species $i$ in the
subvolume, and $N_B$ ($N_{\bar B}$) counts all (anti-)baryons.
In the quark-gluon plasma we have
\begin{eqnarray}
\langle \Delta B^2 \rangle_{\rm QGP} 
  & = & {1\over 9} (N_q + N_{\bar q}),
\nonumber\\
\langle \Delta Q^2 \rangle_{\rm QGP} & \approx &
  {4\over 9} (N_u + N_{\bar u}) + 
  {1\over 9} (N_d + N_{\bar d} + N_s + N_{\bar s}) ,
\label{eq11}
\end{eqnarray}
with the analogous notation of the quark numbers. Neglecting
the contribution from strange quarks, the net charge and baryon
number fluctuations in the QGP are related as
\begin{equation}
\frac{\langle \Delta B^2 \rangle_{\rm QGP}}
  {\langle \Delta Q^2 \rangle_{\rm QGP}} = {2\over 5}\, .
\label{eq12}
\end{equation}

\begin{figure}[htb]
\includegraphics[width=12cm]{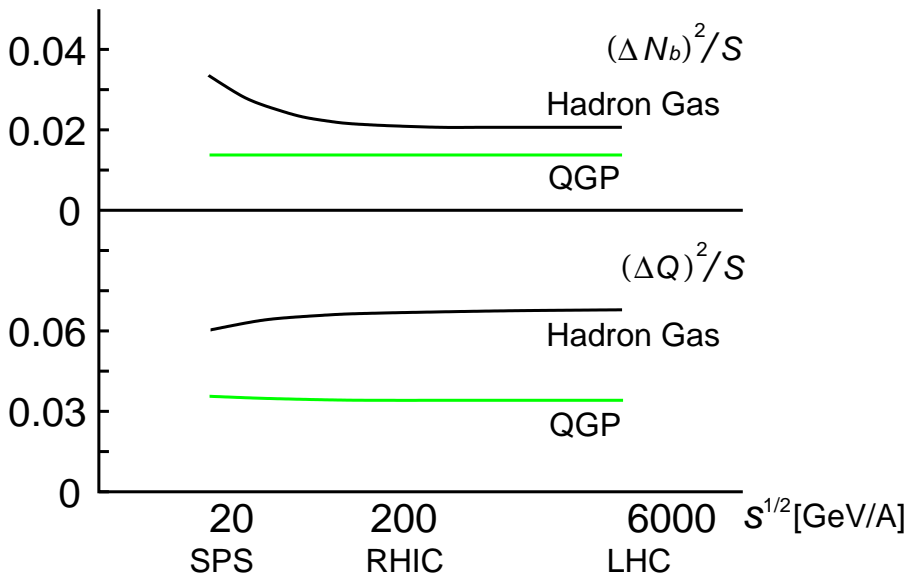}
\caption{Mean square fluctuations per unit entropy of baryon
number and charge for a weakly interacting hadronic gas versus
a weakly interacting quark-gluon plasma. The change with beam
energy is due to the dependence on baryon chemical potential
\protect\cite{AHM00}.}
\label{fig2}
\end{figure}

Since it is difficult, if not impossible, to measure the size of 
an observed subvolume accurately, it is desirable to normalize
the fluctuations to another extensive quantity. One possibility
is to choose the total entropy $S$, which is closely related to
the total number of particles in the subvolume. For the QGP one 
can derive the analytical relation ($\mu$ denotes the baryon 
chemical potential)
\begin{equation}
\frac{\langle \Delta B^2 \rangle_{\rm QGP}}{S_{\rm QGP}}
= \frac{5}{37\pi^2} \left( 1 + {22\over 111}
  \left( {\mu\over\pi T}\right)^2 + \ldots \right) ,
\label{eq13}
\end{equation}
which also determines the charge fluctuations by virtue of 
(\ref{eq12}). Quantitative estimates for the hadron gas phase 
are somewhat more complex, requiring corrections for the
decay of higher resonances. The resulting estimates for the
net charge and baryon number fluctuations, divided by the
entropy, are compared in Fig.~\ref{fig2} with the QGP 
predictions over an energy range spanning from the SPS to the 
LHC. The HG prediction is always significantly higher, and at
RHIC energies and beyond the net charge fluctuations provide
for greater discrimination than the baryon fluctuations. 
However, we will argue shortly that early net baryon number 
fluctuations have a greater chance of survival in a narrow 
rapidity interval.

\begin{figure}[htb]
\includegraphics[width=12cm]{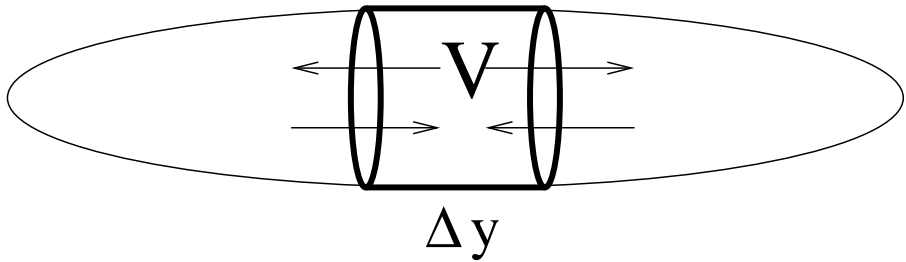}
\caption{We consider net charge and baryon number fluctuations
in a cylindrical volume $V$, defined by a rapidity window
$\Delta y$ in the boost invariant Bjorken model. Conservation
laws dictate that the total charge and baryon number in $V$ can 
only change due to particle transport through the endcaps of $V$.}
\label{fig3}
\end{figure}

Next we address the question whether the fluctuations generated
during a QGP phase can survive the final state expansion in the
HG phase. For simplicity, we consider the boost invariant case
of a longitudinal expansion introduced by Bjorken, illustrated
in Fig.~\ref{fig3}. Considering
a rapidity interval $\Delta y = 1$, we see that the left and
right boundary separate from another with the relative velocity
$\Delta v \approx 0.76 c$. The typical longitudinal velocity 
component of a baryon in the hadronic phase is 
\begin{equation}
{\bar v}_z = (8T/\pi M)^{1/2} \approx 0.32 c .
\label{eq14}
\end{equation}
implying that most baryons will not be able to move out of the
rapidity interval, in which they were contained at the moment of 
hadronization. Thus, we would expect that the baryon number 
fluctuations generated in a rapidity interval $\Delta y \ge 1$
during the QGP phase will remain frozen during the HG phase.
A more detailed analysis of the transport of baryons through
the left- and right-hand surfaces of a cylindrical volume shows
that baryon number fluctuations from the QGP phase can survive
if $\Delta y > {\bar v}_z$ \cite{AHM00}.

The time evolution of fluctuations can also be studied in the
framework of transport theory \cite{SS01}. Consider 
the distribution of particles in rapidity space as a function 
of proper time $\tau$:
\begin{equation}
n(y,\tau) = n_{\rm eq}(y) + f(y,\tau) ,
\label{eq15}
\end{equation}
where $f(y,\tau)$ describes the deviation from equilibrium.
The relaxation toward equilibrium is described by the Langevin 
equation
\begin{equation}
\frac{\partial f}{\partial\tau} = - \frac{1}{\tau_{\rm eq}} f
 + \gamma\frac{\partial^2 f}{\partial y^2} + \xi ,
\label{eq16}
\end{equation}
where $\tau_{\rm eq}$ is the relaxation time, $\gamma$ is the 
coefficient describing diffusion of particles in rapidity,
and$\xi$ is the noise term. 
For a locally conserved particle density, $\tau_{\rm eq}^{-1}=0$,
and the decay of a fluctuation $f(y)$ is solely governed by
diffusion. Making use of the equivalent Fokker-Planck equation for
the Fourier components of the rapidity fluctuations $f(k,\tau)$ 
one can show that the Gaussian width $\sigma_k(\tau)$ of the 
component $f(k,\tau)$ satisfies the equation
\begin{equation}
\frac{\partial}{\partial\tau} \sigma_k^2 
= - 2\gamma k^2 (\sigma_k^2- \chi_k)
\label{eq18}
\end{equation}
where $\chi_k=\int d\eta e^{ik\eta}\langle f(y+\eta)f(y)\rangle$.
There is no relaxation of the fluctuations toward equilibrium 
for the $k=0$ component due to the conservation law; higher 
Fourier components decay exponentially wih the rate $2k^2\gamma$. 

The diffusion constant $\gamma$ is determined by the rapidity 
transfer in particle collisions and the average time between 
collisions: $\gamma = (\delta y_{\rm coll})^2/(2\tau_{\rm coll})$.
With some additional simplifying assumptions one then finds 
that the mean square fluctuation $\langle\Delta N^2\rangle$ 
in a rapidity interval $\Delta y$ decays according to 
\begin{equation}
\langle\Delta N^2\rangle = \Delta y \left( \chi_0 + 
(\sigma_0^2 - \chi_0)G(\Delta y_{\rm diff}/\Delta y)\right) ,
\label{eq21}
\end{equation}
where $G(x)$ is a universal function. For large times, 
$\langle\Delta N^2\rangle$ decays slowly toward its equilibrium 
value.  $\Delta y_{\rm diff}^2 = 2 \int \gamma(\tau) d\tau$
describes the minimal rapidity interval for which initial state 
fluctuations survive.  Shuryak and Stephanov obtained the 
estimates \cite{SS01}
\begin{equation}
\Delta y_{\rm diff}^{(\pi)} \approx 2.2 ,\qquad
\Delta y_{\rm diff}^{(N)} \approx 0.9 
\label{eq24}
\end{equation}
for the rapidity intervals controlling charge and baryon number
fluctuations, respectively. However, these estimates are based on
collision rates deduced from a hadronic cascade model (RQMD) that
does not include a deconfined phase, thus possibly overestimating
the size of the required rapidity windows.

\section{EXPERIMENTAL CONSIDERATIONS AND RESULTS}

For the net charge fluctuations, an elegant way to eliminate the
size of the observed volume is to consider the
fluctuations in the quantity $R=N_+/N_-$. These can be related to
the net charge fluctuations by means of the relation \cite{JK00}
\begin{equation}
D = \langle N_{\rm ch}\rangle\, \langle\Delta R^2\rangle
= 4 \langle\Delta Q^2\rangle / \langle N_{\rm ch}\rangle .
\label{eq25}
\end{equation}
It is useful to correct the quantity $D$ for two trivial effects 
by defining a variable $\tilde D$ as \cite{BJK00}
\begin{equation}
D = C_\mu C_y {\tilde D} ,
\label{eq26}
\end{equation}
with $C_\mu = \langle N_+\rangle^2 / \langle N_-\rangle^2$ 
accounting for the net average charge in the rapidity interval
(due to the baryon excess) and $C_y = 1 - 
\langle N_{\rm ch}\rangle_{\Delta y} / N_{\rm ch}^{(\rm tot)}$ 
correcting for the effect of global charge conservation.
For an uncorrelated thermal pion gas one expects ${\tilde D}=4$. 
Decays of higher meson resonances (such as $\omega, \rho, \eta$)
reduce the value of $\tilde D$ by about 30\%. For a weakly
interacting QGP one anticipates ${\tilde D}\approx 1$.

Preliminary results for $\tilde D$ have been reported by the NA49
collaboration for Pb+Pb collisions at 40, 80, and 158 GeV/u beam
energy \cite{NA49}. A value slightly in excess of ${\tilde D}=4$ 
was found for all three beam energies and independent of the 
selected pseudorapidity interval $0.3 < \Delta\eta < 3.5$. 
The STAR collaboration has reported first results for net
charge fluctuations in Au+Au collisions at $\sqrt{s_{NN}}=130$ 
GeV in the pseudorapidity window $\vert\Delta\eta\vert < 0.7$. 
The value obtained by STAR is ${\tilde D}\approx 3$, as expected
for a nearly baryon-free hadron resonance gas \cite{STAR}.

\section{SUMMARY AND OUTLOOK}

Fluctuations are sensitive probes of the microscopic structure of
dense matter. Local temperature fluctuations can indicate the
presence of a second-order phase transition or critical point, 
while net charge and baryon number fluctuations provide information
about the particle modes that govern the charge and baryon number
transport. In all cases, the rapid expansion of the fireball in its
last stages is essential for the survival of fluctuations established
at early times. 

In addition to thermal fluctuations, the fireball may exhibit
fluctuations due to the initial state \cite{GM00}. These could be
probed independently in p+p or p+A collisions. For example, such
fluctuations may be probes of the quasiclassical coherent glue
fields in a fast-moving nucleus \cite{KLM01}. Similarly, large
nonstatistical fluctuations, such as disoriented chiral condensates,
can signal the presence of unstable collective modes at some time
during the course of the heavy ion collision.

Preliminary data from the NA49 and STAR experiments indicate the 
presence of both, nontrivial $p_T$ fluctuations near central rapidity
and net charge fluctuations similar to those of a pion or hadron
resonance gas. The measurement of net baryon number fluctuations
remains a formidable experimental challenge, since they require the
detection of neutral as well as charged baryons. However, the case
may not be entirely hopeless, as plans by the PHENIX collaboration 
to measure antineutrons with high efficiency demonstrate 
\cite{Ewell}. 

Fluctuations are a rich topic with many theoretical challenges and
opportunities. Lattice gauge theory can make quantitative predictions
for fluctuations near thermal equilibrium. The wide array of possible
fluctuation observables remains largely unexplored. Microscopic models
can simulate final state effects on fluctuations and provide useful
guidance about which observables contain valuable information. There
is ample room for the exploration of novel analysis strategies, as
our picture of the space-time evolution of nuclear collisions at RHIC
energy comes into focus. One especially promising class of observables
are the {\it balance functions}, which can provide a more differential
measure of quantum number fluctuations in phase space \cite{BDP00}.
The subject also promises a wealth of experimental data, as several 
RHIC detectors, foremost PHOBOS and STAR, are well equipped to explore 
event-by-event fluctuations in a wide range of observables.

\end{document}